\documentclass[final,3p,times,twocolumn,UTF8]{elsarticle}

\usepackage{amssymb}
\usepackage{amsthm}
\usepackage{amsmath}
\usepackage{wallpaper}
\usepackage{dcolumn}
\usepackage{bm}
\usepackage{pdfpages}
\usepackage{graphics}
\usepackage[colorlinks,plainpages=false]{hyperref}
\usepackage{bbding}
\usepackage{booktabs}
\usepackage{lscape}
\usepackage{multirow}
\usepackage{color}
\usepackage{ulem}
\usepackage{float}
\usepackage{pdflscape}
\usepackage{afterpage}

\newcommand{\be}{\begin{eqnarray}}
\newcommand{\ee}{\end{eqnarray}}

\journal{Physics Letters B}

\begin{document}
\begin{frontmatter}

\title{Charge symmetry breaking in hypernuclei within RMF model}

\author[SunAdr1,SunAdr2,SunAdr3]{Ting-Ting~Sun}

\author[TanimuraAdr1]{Yusuke Tanimura}

\author[SunAdr2,SagawaAdr1]{Hiroyuki Sagawa\corref{cor1}}
\ead{hiroyuki.sagawa@gmail.com}

\author[HiyamaAdr1,SunAdr2]{Emiko Hiyama\corref{cor1}}
\ead{hiyama@riken.jp}

\cortext[cor1]{Corresponding author.}

\address[SunAdr1]{School of Physics, Zhengzhou University, Zhengzhou 450001, China}
\address[SunAdr2]{RIKEN Nishina Center for Accelerator-Based Science, Wako, Saitama 351-0198, Japan}

\address[TanimuraAdr1]{Department of Physics and Origin of Matter and Evolution of Galaxy (OMEG) Institute, Soongsil University, Seoul 06978, Korea}
\address[SagawaAdr1]{Center for Mathematics and Physics, the University of Aizu, Aizu-Wakamatsu, Fukushima 965-8580, Japan}
\address[HiyamaAdr1]{Department of Physics, Graduate School of Science, Tohoku University, Sendai 980-8578, Japan}

\begin{abstract}
We study the charge symmetry breaking (CSB) effect in the binding energy of mirror hypernuclei in the mass region $A=7\sim 48$ in relativistic mean field (RMF) models introducing $NN$ and $\Lambda N$ interactions. The phenomenological $\Lambda N$ CSB interaction is introduced and the strength parameter is fitted to reproduce the experimental binding energy difference between the mirror hypernuclei $^{12}_\Lambda$B and $^{12}_\Lambda$C. This model is applied to calculate the CSB energy anomaly in mirror hypernuclei with the mass $A=7\sim48$. The model is further applied to predict the binding energy difference of mirror hypernuclei of $A$=40 with the isospin $T=1/2$, $3/2$ and $5/2$ nuclei together with various hyper Ca isotopes and their mirror hypernuclei. Finally the binding energy systematics of $A=$48 hypernuclei are predicted with/without the CSB effect by the PK1 and TM2 energy density functionals (EDFs).
\end{abstract}


\end{frontmatter}

\section{\label{sec:intro}Introduction}

In hypernuclear physics, it is important to extract information on
hyperon ($Y$)-nucleon($N$) interaction. Historically, due to
the difficulties of $YN$ scattering experiments, we have been obtaining
information on $YN$ interaction by the studies of hypernuclear structures.
For this purpose, in the case of the $\Lambda N$ sector, high resolution
$\gamma$-ray experiments have been performed in light $\Lambda$ hypernuclei
systematically~\cite{Tamura-PPNP}.

Theoretically, several shell-model calculations for light $\Lambda$ hypernuclei have been performed \cite{Millner,Umeya}.
Furthermore, microscopic calculations of three- and four-cluster
system with sufficient numerical accuracy have been performed \cite{Korennov,Hiyama1996,Hiyama2009}.
With these theoretical calculations and experimental data,
we have obtained spin-dependent forces of $\Lambda N$ interaction
such as spin-spin term, spin-orbit term, and etc.

However, there is still an open important issue to be solved, i.e.,
charge symmetry breaking (CSB) component in the $\Lambda N$ interaction.
Historically, evidence for the CSB interaction has been observed in the single-$\Lambda$ binding energies $B_\Lambda$ of the $A=4$ mirror hypernuclei $^4_{\Lambda}$H and $^4_{\Lambda}$He. This evidence is attributed to the energy difference $\Delta B_{\Lambda} \equiv B_{\Lambda}(^4_{\Lambda}\text{H})- B_{\Lambda}(^4_{\Lambda}\text{He})$, which was measured to be $-0.35 \pm 0.06$ MeV for the ground $(0^+)$ state and $-0.24 \pm 0.06$ MeV for the excited $(1^+)$ state, respectively \cite{Juric}.
To reproduce the observed data, many theoretical efforts have been done.
To understand the CSB effect, Dalitz and Von Hippel \cite{Dalitz}
pointed out that the $\Lambda$-$\Sigma^0$ mixing mechanism, which is
related to $\Lambda N$-$\Sigma N$ coupling, is important.
Thereafter, several calculations taking account of
$\Lambda N$-$\Sigma N$ coupling have been performed for
the $A=4$ hypernuclei \cite{Nogga_PRL,Haidenbauer,Nogga_NPA,Hiyama}. However, it was difficult to reproduce the experimental data.

In 2015, the ground state of $^4_{\Lambda}$H has been observed with
high accuracy at MAMI-C \cite{MAMI-C}, and the
obtained single-$\Lambda$ separation energy $B_{\Lambda}=2.21 \pm 0.01~({\rm stat})\pm 0.09~({\rm syst})$ MeV, which is consistent with the emulsion value \cite{Juric}. Moreover, the $\gamma$-ray transition from the $1^+$ excited state to the $0^+$ ground state in $^4_{\Lambda}$He has been observed at J-PARC and the obtained excitation energy is $1.406 \pm 0.002~({\rm stat}) \pm 0.002~({\rm syst})$ MeV \cite{Yamamoto}, which was quite different with the old data. After these two observations, many works have discussed
the CSB effect on the $A=4$ light systems \cite{Gal2015,Gal2016,schafer}.
Still, it is difficult to reproduce the new data.

More information on CSB effect are required. For this purpose,
one can also focus on the study of heavier $\Lambda$ hypernuclei with
mass number $A=7, 10$ and larger. For the $A=7$ single-$\Lambda$ hypernuclei, the binding energies of $^7_{\Lambda}$He, $^7_{\Lambda}$Li, and $^7_{\Lambda}$Be with isospin $T=1$, have been observed experimentally. Among these data, the binding energy $B_\Lambda$ in $^7_{\Lambda}$He
has been observed to be $5.68 \pm 0.03~({\rm stat}) \pm 0.25~({\rm syst})$ MeV by the
$^7{\rm Li}(e,e' K^+) ^7_{\Lambda}$He reaction at JLab \cite{JLAB}.
Later, the observed binding energy of $^7_{\Lambda}$He has been updated to be $5.55 \pm 0.10~({\rm stat}) \pm 0.11~({\rm syst})$~MeV with a better systematic error \cite{PRC2016Gogami}.
In the case of $^7_{\Lambda}$Be, there were old emulsion data giving
$B_\Lambda=5.16$~MeV \cite{Juric}. As a result, the energy difference, $\Delta B_\Lambda\equiv B_{\Lambda}(^7_{\Lambda}{\rm He})-
B_{\Lambda}(^7_{\Lambda}{\rm Be})$, between $^7_{\Lambda}$He and
$^7_{\Lambda}$Be is $0.39$~MeV, which exhibits a larger energy difference compared to that of $A=4$ hypernuclei.

Regarding the $A=10$ mirror hypernuclei, there was one old emulsion data for $^{10}_{\Lambda}$Be which gives $B_\Lambda=9.11\pm 0.22$~MeV~\cite{Juric,Cantwell}. In 2016, high-resolution experiment has been done at JLab using the $(e,e'K^+)$ reaction and reported the observed binding energy $B_{\Lambda}$ of $^{10}_{\Lambda}$Be to be $8.60 \pm 0.07~({\rm stat})\pm0.11~({\rm syst})$ MeV \cite{Gogami}.
Meanwhile, the $B_\Lambda$ for $^{10}_\Lambda{\rm B}$ is $8.89 \pm 0.12~({\rm stat})\pm0.04~({\rm syst})$ MeV measured in emulsion~\cite{Davis} while $8.1\pm 0.1$~MeV measured by the $(\pi^+,K^+)$ reaction at KEK~\cite{Hasegawa}, which
was corrected to be $B_{\Lambda}=8.64\pm 0.1$~MeV in Ref.~\cite{Gogami}.
Thus, based on those experimental data, there are two suggested possible binding energy differences $\Delta B_{\Lambda}\equiv B_{\Lambda}(^{10}_\Lambda{\rm Be})-B_{\Lambda}(^{10}_\Lambda{\rm B})$ between the $A=10$ mirror hypernuclei. One value is $\Delta B_\Lambda=9.11-8.89=0.22$~MeV, while the other is $8.60-8.64=-0.04$~MeV.
For more heavier $\Lambda$ hypernuclei, we have data for the $A=12$ mirror hypernuclei $^{12}_{\Lambda}$B and $^{12}_{\Lambda}$C, and the
observed $B_{\Lambda}$s are $11.529 \pm 0.025$ MeV~\cite{TangJLab} and $11.30 \pm 0.19$ MeV~\cite{Gogami,Davis,Dluzewski}, respectively\footnote{According to Mainz compilation \cite{Mainz}, the data is $11.335 \pm 0.126$ MeV, which is similar with Ref. \cite{Gogami}. It should be noted that Ref. \cite{TangJLab} provides high statics. Thus, we cite Ref. \cite{Gogami}.}. Therefore, the experimental value of $\Delta B_{\Lambda}\equiv B_{\Lambda}(^{12}_\Lambda{\rm B})-B_{\Lambda}(^{12}_\Lambda{\rm C})$ is 0.229 MeV, although the error bar in $^{12}_{\Lambda}$C is still large. A comprehensive summary of $B_{\Lambda}$ values for $\Lambda$-hypernuclei with $A \leq 16$ has been provided in Ref.~\cite{Botta}.

It is noted that in mirror hypernuclei, the binding energies $B_{\Lambda}$s on the neutron-rich side are larger compared to those on the proton-rich side for $p$-shell $\Lambda$ hypernuclei. In contrast, the behavior of $s$-shell $\Lambda$ hypernuclei is opposite to that of $p$-shell $\Lambda$ hypernuclei. In order to reproduce both the $s$-shell and $p$-shell hypernuclear data, in Ref.~\cite{Hiyama2009}, one of the present authors, E.H., introduced a phenomenological odd-state CSB interaction which has an opposite sign to the even-state CSB interaction to reproduce the observed data of $^4_{\Lambda}$H and $^4_{\Lambda}$He. The odd-state CSB interaction was adjusted so as to reproduce the observed $B_{\Lambda}$s of $T=1$ isotriplet hypernuclei $^7_{\Lambda}$He, $^7_{\Lambda}$Li, and $^7_{\Lambda}$Be.
The CSB interaction was also applied to calculate the binding energies of the $A=10$ mirror hypernuclei, $^{10}_{\Lambda}$B
and $^{10}_{\Lambda}$Be, and the obtained $\Delta B_{\Lambda}$ was consistent with the data, provided that the odd-state CSB interaction has an opposite sign to that of the even-state CSB interaction.

To further study CSB effect, we need information on heavier $\Lambda$ hypernuclei. For this purpose, it is planned at JLab to produce $^{40}_{\Lambda}$K and $^{48}_{\Lambda}$K via $(e, e'K^+)$ reactions using $^{40}$Ca and $^{48}$Ca targets. In this paper, based on the relativistic mean field (RMF) model, we discuss the CSB effect on the single-$\Lambda$ hypernuclei with mass numbers ranging from $A=7$ to $48$ and also predict the possibility to observe the CSB effect at JLab and J-PARC. In Section~\ref{sec:the}, the theoretical framework
is given. The calculated results and discussions are presented in Section~\ref{sec:results} and
the summary is drawn in Section \ref{sec:summary}.

\section{\label{sec:the}Theoretical framework}
\subsection{RMF model for $\Lambda$ hypernuclei}
RMF models have achieved great successes in the descriptions of ordinary nuclei~\cite{RMFReview}, hypernuclei~\cite{book2016,Mares,Hagino,LV2011,Sun} as well as baryon matter~\cite{Sun2019PRD,Xia2023PRD}. The starting point of the meson-exchange RMF model for the $\Lambda$ hypernuclei is the following covariant Lagrangian density,
\begin{equation}
\mathcal{L}=\mathcal{L}_{N}+\mathcal{L}_{\Lambda}.
\end{equation}
Here $\mathcal{L}_{N}$ is the standard RMF Lagrangian density for nucleons~\cite{RMFReview}, in which the couplings with the scalar-isoscalar $\sigma$, vector-isoscalar $\omega_{\mu}$, and vector-isovector $\vec{\rho}_{\mu}$ mesons, and the photon $A_{\mu}$ are included, i.e.,
\begin{align}
\mathcal{L}_{N}=&\sum_{i=n,p}\bar{\psi}_i\left[i\gamma^{\mu}\partial_{\mu}-M_i-g_{\sigma i}\sigma-g_{\omega i}\gamma^\mu\omega_\mu\right.\nonumber\\
&\left.-g_{\rho i}\gamma^{\mu}\vec{\tau}_{i}\cdot\vec{\rho}_{\mu}-e\gamma^{\mu}A_{\mu}\frac{1-\tau_{i,3}}{2}
\right]\psi_i\nonumber\\
&+\frac{1}{2}\partial_{\mu}\sigma\partial^\mu\sigma-\frac{1}{2}m_{\sigma}^{2}\sigma^2-\frac{1}{3}g_2\sigma^3-\frac{1}{4}g_3\sigma^4\nonumber\\
&-\frac{1}{4}\Omega_{\mu\nu}\Omega^{\mu\nu}+\frac{1}{2}m_{\omega}^2\omega_{\mu}\omega^\mu+
\frac{1}{4}c_3(\omega_\mu\omega^\mu)^2\nonumber\\
&-\frac{1}{4}\vec{R}_{\mu\nu}\cdot\vec{R}^{\mu\nu}+\frac{1}{2}m_{\rho}^2\vec{\rho}_{\mu}\cdot
\vec{\rho}^{\mu}-\frac{1}{4}F_{\mu\nu}F^{\mu\nu},
\end{align}
where $M_i~(i=n,p)$ denotes the nucleon mass, $\vec{\tau}_i$ is the isospin with the $3$rd component $\tau_{i,3}$ ($+1$ for neutrons and $-1$ for protons), and $m_\phi~(\phi=\sigma, \omega, \rho)$ and $g_{\phi i}$ are the masses and coupling constants for mesons, respectively.
$\Omega_{\mu\nu}$, $\vec{R}_{\mu\nu}$, and $F_{\mu\nu}$ are the field tensors for the $\omega$ and $\vec{\rho}$ mesons and photons. $g_2$, $g_3$, and $c_3$ are the parameters introduced in the nonlinear self-coupling terms.

The Lagrangian density $\mathcal{L}_{\Lambda}$ represents the contributions from $\Lambda$ hyperons, in which only the couplings with the $\sigma$ and $\omega_{\mu}$ mesons are included because of $\Lambda$ hyperons being charge neutral and zero isospin, i.e.,
\begin{align}
\mathcal{L}_{\Lambda}=&\bar{\psi}_{\Lambda}\left[i\gamma^{\mu}
\partial_{\mu}-M_{\Lambda}-g_{\sigma\Lambda}\sigma-g_{\omega\Lambda}
\gamma^{\mu}\omega_{\mu}\right.\nonumber\\
&\left.-\frac{f_{\omega\Lambda\Lambda}}{2m_{\Lambda}}\sigma^{\mu\nu}\partial_{\nu}\omega_{\mu}\right]\psi_{\Lambda},
\end{align}
where $M_\Lambda$ is the mass of the $\Lambda$ hyperon, $g_{\sigma\Lambda}$ and $g_{\omega\Lambda}$ are the coupling constants with the $\sigma$ and $\omega$ meson, respectively. To reproduce the small $\Lambda$ spin-orbit splitting, a term of $\omega\Lambda\Lambda$ tensor coupling is introduced with $f_{\omega\Lambda\Lambda}$ being the coupling constant.

For a system with time-reversal symmetry, the space-like
components of the vector fields $\omega_{\mu}$ and $\vec{\rho}_{\mu}$ vanish, leaving only the time components $\omega_0$ and $\vec{\rho}_0$. Meanwhile, charge
conservation guarantees that only the third component $\rho_{0,3}$ in the isospin space of $\vec{\rho}_0$ exists. With the mean-field and no-sea approximations, the single-particle Dirac equations for nucleons and hyperons and the Klein-Gordon equations for mesons and photons can be obtained by the variational procedure.

With spherical symmetry, the Dirac spinor for nucleons and hyperons can be expanded as
\begin{equation}
\psi_{n\kappa m}({\bm r})=\frac{1}{r}
                                     \left(
                                            \begin{array}{c}
                                              iG_{n\kappa}(r)\\
                                              -F_{n\kappa}(r){\bm \sigma}\cdot\hat{\bm r} \\
                                            \end{array}
                                          \right)Y_{jm}^{l}(\theta,\phi),
\end{equation}
where $G_{n\kappa}(r)/r$ and $F_{n\kappa}(r)/r$ are the radial wave functions for the upper and lower components, $Y_{jm}^{l}(\theta,\phi)$ is the spinor spherical harmonic. The quantum number $\kappa$ is defined as $\kappa=(-1)^{j+l+1/2}(j+1/2)$. With the radial wave functions, the radial Dirac equations for baryons $(i=n,p,\Lambda)$ can be obtained as,
\begin{equation}
\left(
  \begin{array}{cc}
    V_i+S_i & {\displaystyle -\frac{d}{dr}+\frac{\kappa}{r}+T_i} \\
    {\displaystyle \frac{d}{dr}+\frac{\kappa}{r}+T_i} & V_i-S_i-2M_{i} \\
  \end{array}
\right)
 \left(
   \begin{array}{c}
     G_{n\kappa} \\
     F_{n\kappa} \\
   \end{array}
 \right)
 =\varepsilon_{n\kappa}
  \left(
   \begin{array}{c}
     G_{n\kappa} \\
     F_{n\kappa} \\
   \end{array}
 \right),
\end{equation}
where $S_i$, $V_i$, and $T_i$ are respectively the mean-field scalar potential, vector potential, and the $\omega \Lambda\Lambda$ tensor potential, 
\begin{subequations}
\begin{align}
&S_i=g_{\sigma i}\sigma,\label{Eq:Si}\\
&V_i=g_{\omega i}\omega_0+g_{\rho i}\tau_{i,3}\rho_{0,3}+\frac{1}{2}e(1-\tau_{i,3})A_0,
\label{Eq:Vi}\\
&T_i=-\frac{f_{\omega \Lambda\Lambda}}{2M_i}\partial_r\omega_0.
\label{Eq:Ti}
\end{align}
\end{subequations}
Note that the terms related to $\rho_{0,3}$ and $A_{0}$ in Eq.~(
\ref{Eq:Vi}) are zero for $\Lambda$ hyperons, while the tensor potential in Eq.~(\ref{Eq:Ti}) is zero for nucleons. The $\omega\Lambda\Lambda$ tensor interaction was introduced to reproduce the experimentally observed small spin-orbit splitting for $\Lambda$ hyperon~\cite{Jennings,Cohen}.

The Klein-Gordon equations for mesons and photons are
\begin{equation}
(\partial^\mu\partial_\mu+m_{\phi}^{2})\phi=S_\phi,
\end{equation}
with the source terms
\begin{equation}
S_{\phi}=
\left\{
  \begin{array}{ll}
    \sum\limits_{ i=n,p,\Lambda}-g_{\sigma i}\rho_{s i}-g_{2}\sigma^2-g_3 \sigma^3, & \hbox{$\phi=\sigma$;} \\
    \sum\limits_{i=n,p,\Lambda} g_{\omega i}\rho_{v i}+\frac{f_{\omega\Lambda\Lambda}}{2M_{\Lambda}}\partial_k j^{0k}_{T\Lambda}-c_3 \omega_0^3, & \hbox{$\phi=\omega$;} \\
    \sum\limits_{i=n,p}g_{\rho i}\tau_{i,3}\rho_{v i}, & \hbox{$\phi=\rho$;} \\
    e\rho_c, & \hbox{$\phi=A$;}
  \end{array}
\right.
\label{Eq:KG}
\end{equation}
where $\rho_{s i}$ and $\rho_{v i}$ are the scalar and vector densities for nucleons and hyperons, $j_{T\Lambda}^{0k}$ is the tensor density for $\Lambda$ hyperons, and $\rho_c$ is the charge density for protons.

With the radial wave functions, those densities in Eq.~(\ref{Eq:KG}) can be obtained as
\begin{subequations}
\begin{align}
\rho_{si}(r)&=\frac{1}{4\pi r^2}\sum\limits_{k=1}^{A_i}\left[G^2_{k i}(r)-F_{ki}^{2}(r)\right],\\
\rho_{vi}(r)&=\frac{1}{4\pi r^2}\sum\limits_{k=1}^{A_i}\left[G^2_{k i}(r)+F_{ki}^{2}(r)\right],\\
\rho_{c}(r)&=\frac{1}{4\pi r^2}\sum\limits_{k=1}^{A_p}\left[G^2_{k p}(r)+F_{kp}^{2}(r)\right],\\
{\bm j}_{T\Lambda}^0(r)&=\frac{1}{4\pi r^2}\sum\limits_{k=1}^{A_\Lambda}\left[2G_{k \Lambda}(r)F_{k\Lambda}(r)\right]{\bm n},
\label{Eq:rho3}
\end{align}
\end{subequations}
where ${\bm n}$ is the angular unit vector. The baryon number $A_{i}~(i=n,p,\Lambda)$ can be calculated by integrating the baryon density $\rho_{vi}(r)$ in coordinate space as
\begin{equation}
A_i=\int 4\pi r^2 dr\rho_{vi}(r).
\end{equation}

\subsection{$\Lambda N$ CSB interaction}
In analogy of non-relativistic CSB interaction~\cite{Sagawa}, we introduce a simple $\Lambda N$ charge symmetry breaking (CSB) interaction as
\be  \label{LN-CSB}
V^{\Lambda N}_{CSB}=
\frac{1}{2}V_0^{\Lambda N}\sum_{k=1}^{A}\bar{\psi}_\Lambda\gamma_\mu\psi_\Lambda \bar{\psi}_{k}\gamma^\mu \tau_{3}\psi_{k},
\ee
where $V_0^{\Lambda N}$ is the strength of CSB interaction, $\tau_{3}$ is the $3$rd component of isospin for neutrons or protons.
The energy density functional is obtained for the interaction \eqref{LN-CSB} as,
\begin{equation}
\varepsilon_{\Lambda N}= \frac{1}{2}V_0^{\Lambda N}\rho_{\Lambda}(\rho_n-\rho_p),
\end{equation}
where $\rho_{n}(r)$  and $\rho_{p}(r)$ are the baryon densities of neutron and proton, respectively. With this CSB interaction, the attractive $\Lambda n$ interaction is strengthen while $\Lambda p$ interaction is weaken for a negative $V_0^{\Lambda N}$ value.

With or without the $\Lambda N$ CSB interaction, the Dirac equations for baryons ($n$, $p$, $\Lambda$), the Klein-Gordon equations for mesons and photon, the mean-field potentials, and densities in the RMF model are solved by iteration procedure in the coordinate space. The single-$\Lambda$ binding energy $B_{\Lambda}$ is calculated by using a formula,
\begin{equation}
B_{\Lambda}(Z, N, 1)=B(^{A}_{\Lambda}Z)-B(^{A-1}Z),
\end{equation}
where $B(^{A}_{\Lambda}Z)$ and $B(^{A-1}Z)$ are the binding energies for the single-$\Lambda$ hypernuclei and the corresponding core nuclei, respectively.  
The difference in single-$\Lambda$ binding energies between mirror hypernuclei is then defined for $N \geq Z$ as
\begin{equation}
\Delta B_\Lambda(A=N+Z)=B_\Lambda(Z,N,1)-B_\Lambda(N,Z,1). 
\end{equation}

Equations in the RMF model are solved in the coordinate space with a size of $R=20$ fm and a step size of $\Delta r=0.05$ fm. For the $NN$ interaction, the PK1~\cite{PK1} and TM2~\cite{TM2} parameter sets are adopted. For the $\Lambda N$ interactions, the parameters are listed in Table~\ref{tab:RMF-para}. With those $NN$ and $\Lambda N$ interactions, the single-$\Lambda$ binding energy in $^{12}_{\Lambda}$C~\cite{Gal2016review} and the $\Lambda$ $1p$ spin-orbit splitting in $^{13}_{\Lambda}$C~\cite{Ajimura} are well reproduced as $B_\Lambda$=11.3 MeV and $E(1/2^-)-E(3/2^-)$=0.152 MeV, respectively. The single-$\Lambda$ binding energy in the heavier hypernucleus $^{40}_{\Lambda}$Ca can also be well given, and it is predicted to be 19.24 and 18.45 MeV with PK1 and TM2 EDFs, respectively, while the experimental value is reported as 18.7 $\pm$ 1.1 MeV~\cite{Pile}. The parameter $V_0^{\Lambda N}$ in the CSB interaction is determined to reproduce the $\Lambda$ binding energy difference of the $A=12$ mirror hypernuclei $^{12}_{\Lambda}$B and $^{12}_{\Lambda}$C, i.e., $\Delta B_{\Lambda}=0.229~$MeV given by the experimental values of $B_{\Lambda}=11.529\pm 0.025$~MeV for $^{12}_{\Lambda}$B ~\cite{TangJLab} and $11.30\pm 0.19$~MeV for $^{12}_{\Lambda}$C~\cite{Davis}, both in their ground states. It is noted that the error bar of $B_\Lambda$ for 
$^{12}_{\Lambda}$C is much larger than that for $^{12}_{\Lambda}$B. Here we focus on the centroid values of $B_\Lambda$s in these two hypernuclei with CSB effect. To reproduce this value, $V_0^{\Lambda N}$ is determined as given in Table \ref{tab:RMF-para}. The upper and lower limits of the $\Lambda N$ CSB strength are also evaluated by considering the experimental uncertainties of binding energies.
\begin{table}
\centering
\caption{\label{tab:RMF-para} Parameters of the $\Lambda N$ interactions in the RMF model.}
\begin{tabular}{crr}
\toprule
  &  PK1  &  TM2 \\ \hline
  $g_{\sigma \Lambda}/g_{\sigma N}$  & 0.6206 &  0.6233  \\
  $g_{\omega \Lambda}/g_{\omega N}$  & 0.6666 & 0.6666  \\
   $f_{\omega \Lambda\Lambda}/f_{\omega N}$   & -1.1174 & -1.1210  \\ \hline
   $V_0^{\Lambda N}$ (MeV fm$^3$) &  -33.05 &  -38.00 \\
     upper limit   &  -63.15      &  -72.00   \\
     lower limit   &   -2.85      & -3.70   \\
  \bottomrule
\end{tabular}
\end{table}

\section{\label{sec:results}Results and discussions}

\begin{figure}[t!]
\centering
\includegraphics[width=0.95\linewidth]{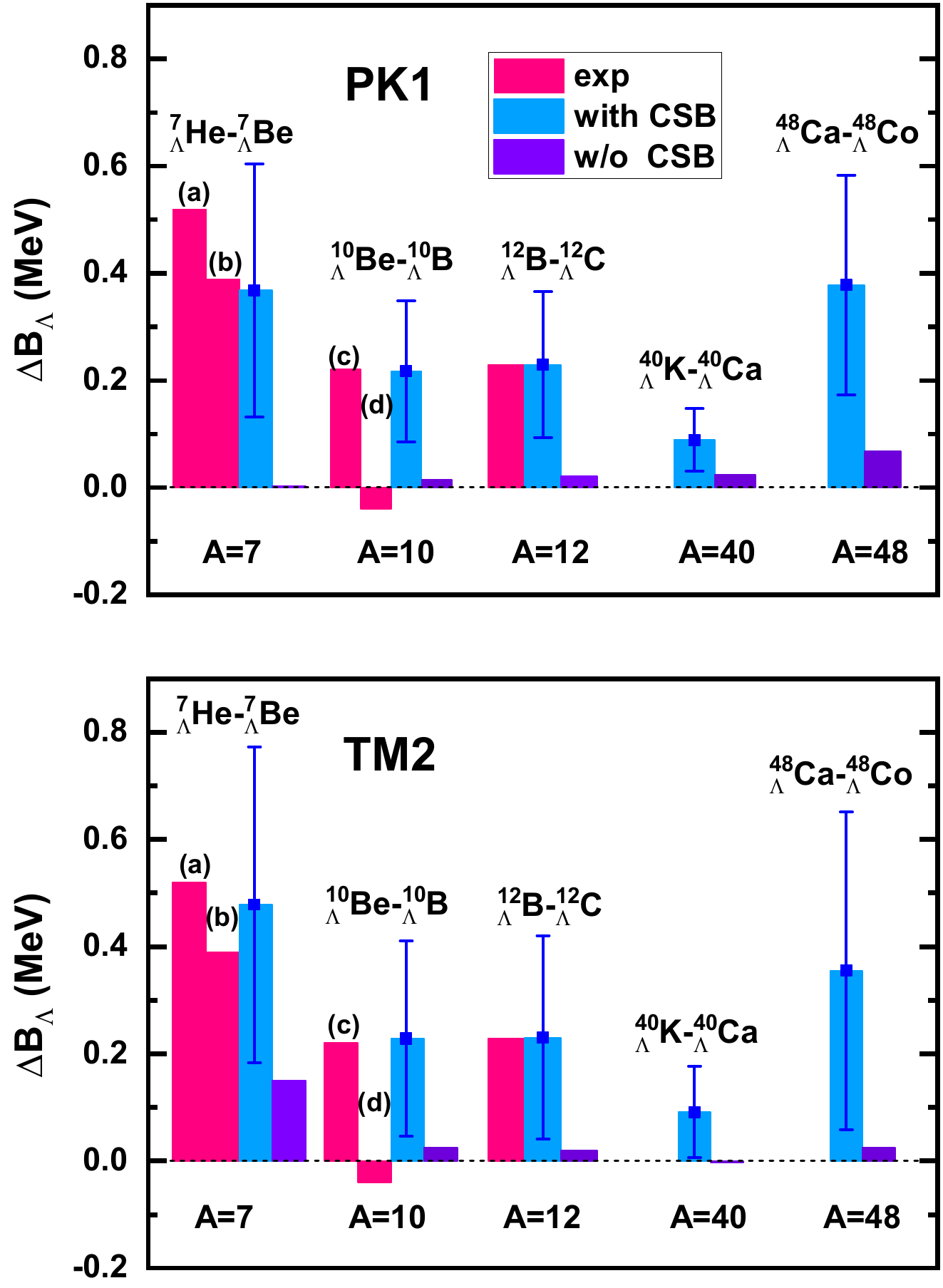}
\caption{\label{Fig1:DBLam_mirrornuclei} The differences  $\Delta B_{\Lambda}$ of the single-$\Lambda$ binding energy between the mirror hypernuclei ($^{7}_{\Lambda}$He, $^{7}_{\Lambda}$Be), ($^{10}_{\Lambda}$Be, $^{10}_{\Lambda}$B), ($^{12}_{\Lambda}$B, $^{12}_{\Lambda}$C), ($^{40}_{\Lambda}$K, $^{40}_{\Lambda}$Ca),
and ($^{48}_{\Lambda}$Ca, $^{48}_{\Lambda}$Co) obtained by the RMF models with and without the $\Lambda N$ CSB interaction, in comparison with the experimental data. The upper (lower) panel shows results with PK1(TM2) EDF.
In $^{10}_{\Lambda}$Be-$^{10}_{\Lambda}$B, 
we cite two experimental data specified in (a) and (b).}
\end{figure}

Using CSB $\Lambda N$ interaction, we first examine the lighter mirror $\Lambda$ hypernuclei with $A=7$ and $A=10$, such as ($^7_{\Lambda}$He, $^7_{\Lambda}$Be) and ($^{10}_{\Lambda}$Be, $^{10}_{\Lambda}$B), as shown in Fig.~\ref{Fig1:DBLam_mirrornuclei}. Among these, $^7_{\Lambda}$He and $^7_{\Lambda}$Be are the lightest $p$-shell $\Lambda$ hypernuclei for the discussions on CSB effect. The $^7_{\Lambda}$Be was observed by emulsion and the binding energy in the ground state is $5.16 \pm 0.08$ MeV \cite{Juric}. The binding energy $B_{\Lambda}$ of $^7_{\Lambda}$He in the ground state has been measured to be $5.55 \pm 0.10 ~({\rm stat}) \pm 0.11$~(syst) MeV~\cite{PRC2016Gogami} at JLab. Thus, $\Delta B_{\Lambda} = B_{\Lambda}(^7_{\Lambda}{\rm He})-B_{\Lambda}(^7_{\Lambda}{\rm Be})=0.39$ MeV. As shown in Fig.~\ref{Fig1:DBLam_mirrornuclei}, our calculated $\Delta B_{\Lambda}$ is about 0.37 MeV by PK1 EDF, which is close to the data of $0.39$~MeV when the $\Lambda N$ interaction is fitted to reproduce the centroid data of the $\Delta B_\Lambda$ for $A=12$ mirror $\Lambda$ hypernuclei. We also fit the $\Lambda N$ interaction strengths to reproduce the binding energy differences across the range of observed values. These results are shown by the blue-colored bars. Our results for $A=7$ hypernuclei are consistent with the data within one $\sigma$ deviation. Especially, it is in good agreement with the data if we use PK1 EDF. For $A=10$ mirror hypernuclei, our calculated $\Delta B_{\Lambda}$ is about 0.2 MeV by both PK1 and TM2 EDFs. As previously mentioned, there are two interpretations of the experimental data for $A=10$ $\Lambda$ hypernuclei. One interpretation, based on Refs.~\cite{Gogami,Davis,Hasegawa}, yields a value of 0.22 MeV, which is consistent with our results (see Fig.~\ref{Fig1:DBLam_mirrornuclei}(a)). The other interpretation, corresponding to Fig.~\ref{Fig1:DBLam_mirrornuclei}(b), gives a value of -0.04 MeV, which is inconsistent with our findings. Due to these conflicting interpretations, it is challenging to give definitive conclusions on the CSB effect in $A=10$ $\Lambda$ hypernuclei. Therefore, further high-resolution experimental data are needed. In fact, it is already planned to measure the binding energy of $^{10}_\Lambda{\rm B}$ with improved accuracy in the J-PARC E94 experiment~\cite{EPJ_Gogami}. 

Next, based on these results, we discuss the $\Delta B_{\Lambda}$ values for mirror $\Lambda$ hypernuclei with $A=40$ and 48. For the combination of $^{40}_{\Lambda}$K and $^{40}_{\Lambda}$Ca, Fig.~\ref{Fig1:DBLam_mirrornuclei} shows that $\Delta B_{\Lambda} \sim 0$ without CSB. When CSB interaction is included, the calculated $\Delta B_{\Lambda}=0.1$ MeV. For $^{40}_{\Lambda}$K, at JLab, it is planned to produce this hypernucleus by the $(e,e'K^+)$ reaction using a $^{40}$Ca target and measure the binding energy of the ground state with a resolution of $100$ keV. Meanwhile, it is possible to produce $^{40}_{\Lambda}$Ca by the $(\pi^+,K^+)$ reaction at J-PARC using a $^{40}$Ca target. Thus, it would be possible to see the CSB effect by these experiments.

Moreover, at JLab, they plan to use a $^{48}$Ca target to produce $^{48}_{\Lambda}$K by $(e,e'K^+)$ reaction. The calculated binding energy of $^{48}_{\Lambda}$K is 21.08 (20.68) MeV for PK1 EDF with (without) $\Lambda N$ CSB interaction, from which we can also see the effect of CSB.
We also calculate the binding energy difference between $A=48$ mirror hypernuclei $^{48}_{\Lambda}$Ca and $^{48}_{\Lambda}$Co to be 0.378 (0.068) MeV with (without) CSB. We see a more pronounced CSB effect compared to the case of $A=40$ hypernuclei. At J-PARC, by $(\pi^+,K^+)$ reaction with a $^{48}$Ca target, $^{48}_{\Lambda}$Ca can be produced and the large CSB effect of 0.378 MeV shown in Fig.~\ref{Fig1:DBLam_mirrornuclei} might be proved in comparison with the energy of $^{48}_{\Lambda}$Co. However, it would be difficult to produce $^{48}_{\Lambda}$Co due to the lack of an appropriate target. From this fact, it would be better to see $\Delta B_\Lambda$ in the mirror hypernuclei $^{40}_{\Lambda}$K and $^{40}_{\Lambda}$Ca, using the $^{40}$Ca target by $(e,e'K^+)$ reaction at JLab and by $(\pi^+,K^+)$ reaction at J-PARC.

\begin{figure}[t!]
\centering
\includegraphics[width=0.95\linewidth]{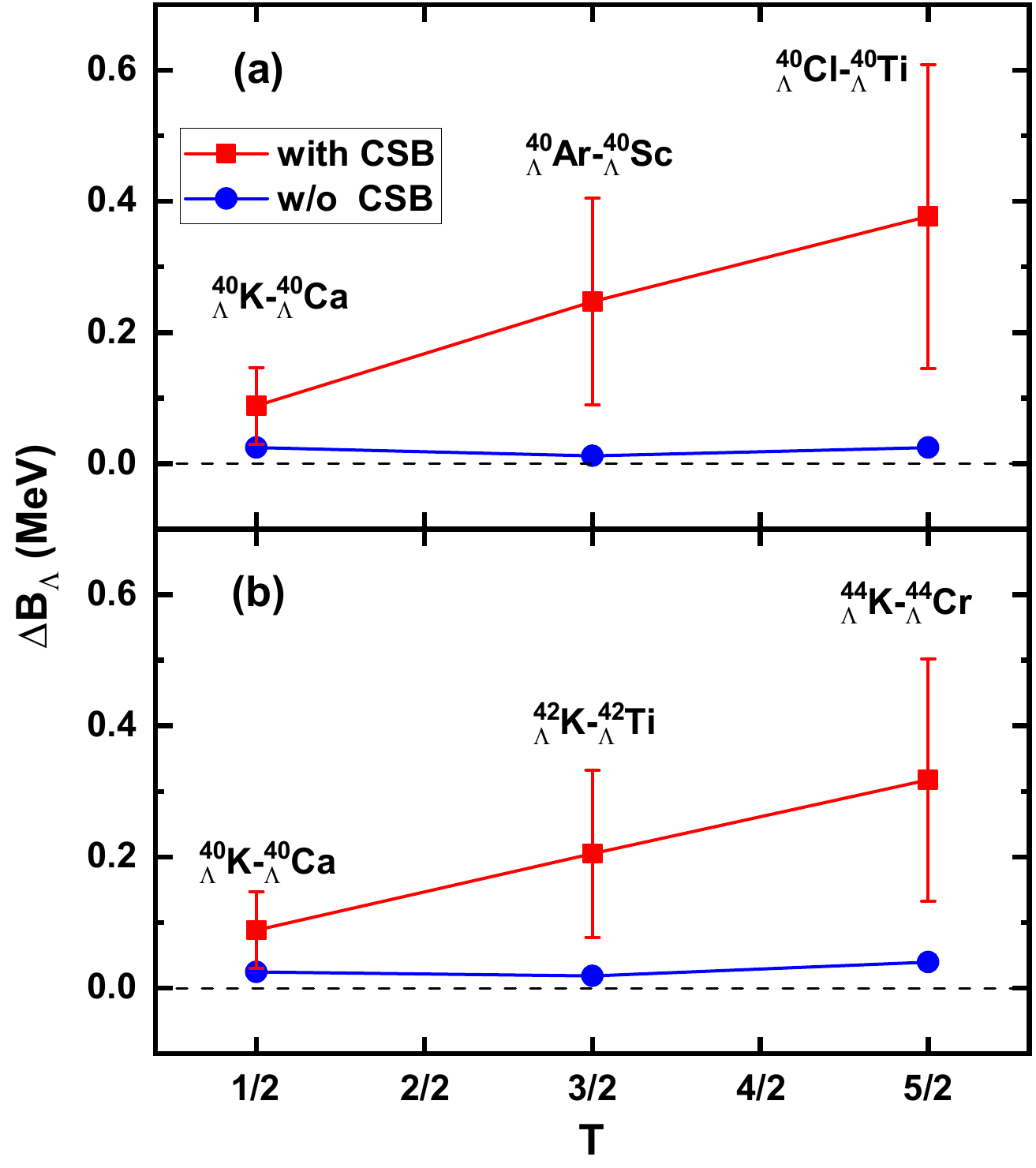}
\caption{\label{Fig2:DBLam_A40} The single-$\Lambda$ binding energy differences $\Delta B_\Lambda$ between the mirror hypernuclei with $A=40$ and those with different isospins, i.e., $T=1/2$ multiplet ($^{40}_{\Lambda}$K, $^{40}_{\Lambda}$Ca), $T=3/2$ multiplets ($^{40}_{\Lambda}$Ar, $^{40}_{\Lambda}$Sc) and ($^{42}_{\Lambda}$K, $^{42}_{\Lambda}$Ti), and $T=5/2$ multiplets ($^{40}_{\Lambda}$Cl, $^{40}_{\Lambda}$Ti) and
($^{44}_{\Lambda}$K, $^{44}_{\Lambda}$Cr), obtained by the PK1 EDF with and without $\Lambda N$ CSB interaction. }
\end{figure}

Recently, the importance of the spin dependent hyperon-nucleon CSB interaction was suggested in light hypernuclear systems in Refs.~\cite{schafer,FB_Haidenbauer,HLe}. For instances, Ref.~\cite{HLe} demonstrated that the effects of this spin dependence can vary significantly, even in sign, for $A=7$ and $A=8$ hypernuclei. In this work, we also examine the impact of spin-dependent CSB by introducing the following spin-spin interaction and spin-spin CSB interaction,
\begin{eqnarray}
&&V_{\sigma\sigma}^{N\Lambda}=g_{N\Lambda}\vec{\sigma}_{N}\cdot\vec{\sigma}_{\Lambda},\\
&&V_{N\Lambda}^{\rm CSB}=g_{N\Lambda}^{\rm CSB}\vec{\sigma}_{N}\cdot\vec{\sigma}_{\Lambda}\frac{1}{2}(\tau_{z}^{N}+\tau_{z}^{\Lambda}),
\end{eqnarray}
where the interaction strength parameters $g_{N\Lambda}$ and $g_{N\Lambda}^{\rm CSB}$ are fixed to reproduce the spin-doublet states in $A=12$ mirror $\Lambda$ hypernuclei~\cite{TangJLab}. The observed energy splitting between the $1^-$ and $2^-$ spin-doublet states is 0.162 MeV in $^{12}_\Lambda$C and 0.179 MeV in $^{12}_\Lambda$B, resulting in a small contribution from the spin-spin CSB interaction of only $0.017$ MeV. Using the spin-spin and spin-spin CSB interactions, we further predict the energy splittings for the ground state doublets in $^{40}_\Lambda$K-$^{40}_\Lambda$Ca, and $^{48}_\Lambda$Ca-$^{48}_\Lambda$Co, respectively. The calculated energy splitting for $^{40}_\Lambda$K and $^{40}_\Lambda$Ca are 0.1074MeV and 0.0972MeV, respectively, while those for $^{48}_\Lambda$Ca and $^{48}_\Lambda$Co are 0.1389 MeV and 0.1534 MeV, respectively. Then the contributions from the spin-spin CSB term are finally obtained to be 0.0102 for $A=40$ and -0.0145 MeV for $A=48$, which are negligibly small compared to the obtained $\Delta B_\Lambda$ values of $0.089\pm0.059$ MeV and $0.378\pm0.205$ MeV, respectively.

Second, let us explain the reason why CSB effect of ($^{40}_{\Lambda}$Ca, $^{40}_{\Lambda}$K) is rather small, i.e., $\Delta B_{\Lambda}$=0.09 MeV.
It is noted that the isospin of ($^{40}_{\Lambda}$Ca, $^{40}_{\Lambda}$K)
is $T=1/2$. Generally speaking, it is better to see $\Delta B_{\Lambda}$
for study of CSB in systems with larger total isospins.
In Fig.~\ref{Fig2:DBLam_A40}, we show the calculated $\Delta B_{\Lambda}$
for ($^{40}_{\Lambda}$Ar, $^{40}_{\Lambda}$Sc) with $T=3/2$ and
for ($^{40}_{\Lambda}$Cl, $^{40}_{\Lambda}$Ti) with $T=5/2$.
We find that the calculated $\Delta B_{\Lambda}$
becomes larger with increasing $T$, that is, $\Delta B_{\Lambda} \sim 0.25$ MeV
for ($^{40}_{\Lambda}$Ar, $^{40}_{\Lambda}$Sc)
and $\Delta B_{\Lambda}\sim 0.4$ MeV for ($^{40}_{\Lambda}$Cl, $^{40}_{\Lambda}$Ti) with $T=5/2$, as increasing total isospin $T$.
To produce the mirror hypernuclei ($^{40}_{\Lambda}$Ar, $^{40}_{\Lambda}$Sc)
and $(^{40}_{\Lambda}${\rm Cl} and $^{40}_{\Lambda}$Ti), 
$^{40}$Ar, $^{40}$Sc, $^{40}$Cl, and $^{40}$Ti target can be used by $(\pi^+,K^+)$ reaction and for the production of $^{40}_\Lambda$Cl and $^{40}_{\Lambda}$Sc, we also can use $^{40}$Ar and $^{40}$Ti targets by $(e,e'K^+)$ reaction at JLab.

As another way to see CSB effect, we propose to use Ca target. Because, several Ca targets such as $^{40}$Ca, $^{42}$Ca,
$^{44}$Ca, $^{46}$Ca, and $^{48}$Ca are available for experiments to produce hypernuclei. By $(\pi^+, K^+)$ reaction, $^{40}_{\Lambda}$Ca,
$^{42}_{\Lambda}$Ca, $^{44}_{\Lambda}$Ca, $^{46}_{\Lambda}$Ca, and $^{48}_{\Lambda}$Ca can be produced.

\begin{figure}[t!]
\centering
\includegraphics[width=0.95\linewidth]{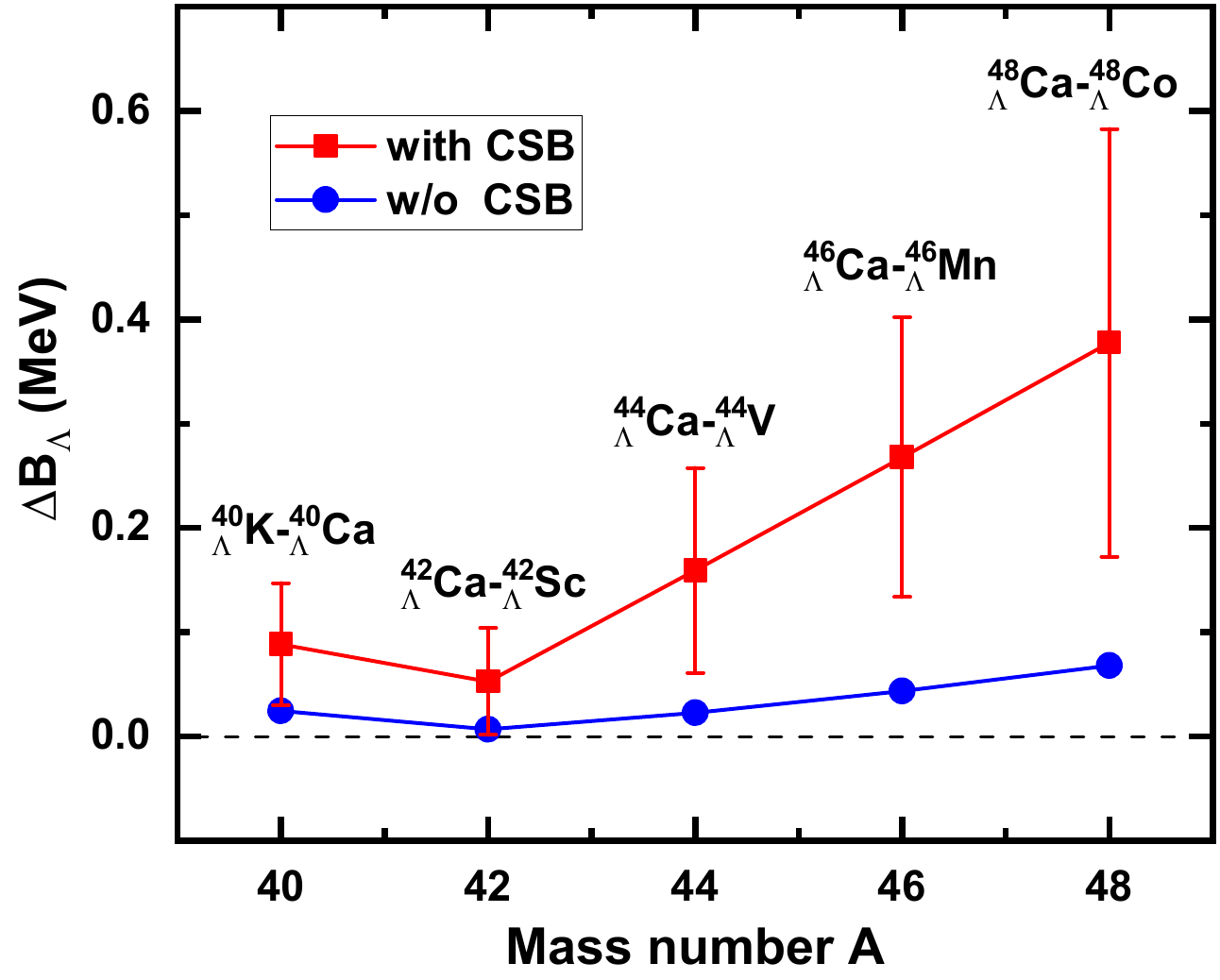}
\caption{\label{Fig3:DBLam_Ca} The single-$\Lambda$ binding energy differences $\Delta B_\Lambda$ between the mirror hypernuclei with different mass number, i.e., ($^{40}_{\Lambda}$K, $^{40}_{\Lambda}$Ca), ($^{42}_{\Lambda}$Ca, $^{42}_{\Lambda}$Sc),
($^{44}_{\Lambda}$Ca, $^{44}_{\Lambda}$V), ($^{46}_{\Lambda}$Ca, $^{46}_{\Lambda}$Mn), ($^{48}_{\Lambda}$Ca, $^{48}_{\Lambda}$Co),
obtained with the PK1 EDF with and without $\Lambda N$ CSB interaction.}
\end{figure}

\begin{figure}[t!]
\centering
\includegraphics[width=0.95\linewidth]{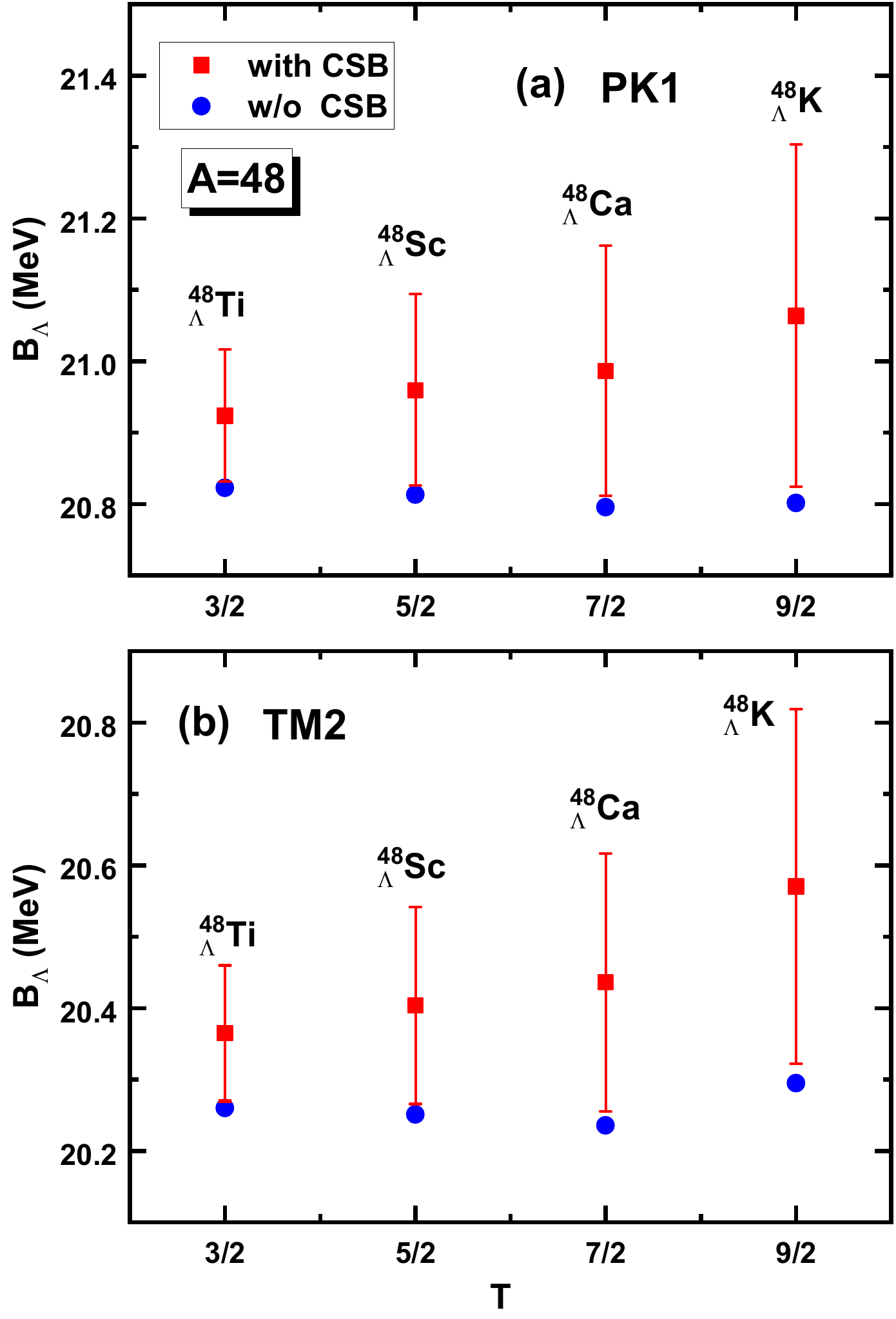}
\caption{\label{Fig4:BLam_A48} The single-$\Lambda$ binding energies $B_{\Lambda}$ for the $A=48$ hypernuclei, i.e., $^{48}_{\Lambda}$Ti, $^{48}_{\Lambda}$Sc, $^{48}_{\Lambda}$Ca, $^{48}_{\Lambda}$K with isospin $T=3/2, 5/2, 7/2, 9/2$, obtained by the RMF model with and without $\Lambda N$ CSB interaction. The upper (lower) panel shows results with PK1(TM2) EDF.}
\end{figure}

In Fig.~\ref{Fig3:DBLam_Ca}, we show the calculated $\Delta B_{\Lambda}$ between hyper Ca isotopes and the corresponding mirror hypernuclei.
The calculated $\Delta B_{\Lambda}$ becomes larger
with larger neutron number. For $A=48$, the calculated $\Delta B_{\Lambda}$ is about 0.4 MeV with CSB effect, which is significantly large.
If we could obtain the observed binding energies of
$^{48}_{\Lambda}$Ca and $^{48}_{\Lambda}$Co, it is useful for the
study of CSB effect in $\Lambda N$ interaction. However, it is difficult to produce $^{48}_{\Lambda}$Co due to the lack of an appropriate target.

In this way, we understand that the larger CSB effect is associated with nuclei with  larger isospins, that is, it is better to see the binding energies
of $A=48$ hypernuclei if there are some target.
Along this line, let us examine nuclei with $A=48$ isobars.
Using $^{48}$Ti, $^{48}$Sc, $^{48}$Ca, and $^{48}$K as targets,
we could produce $^{48}_{\Lambda}$Ti, $^{48}_{\Lambda}$Sc, $^{48}_{\Lambda}$Ca, and $^{48}_{\Lambda}$K by $(\pi^+,K^+)$ reaction.
Figure~\ref{Fig4:BLam_A48} shows the binding energies of the ground states of these $\Lambda$ hypernuclei with/without CSB $\Lambda N$ interaction.
We see that calculated $B_{\Lambda}$ without CSB interaction for these hypernuclei are about 20.8 MeV with PK1 EDF in Fig.~\ref{Fig4:BLam_A48}(a), and the isospin dependence is small. This is also the case in the TM2 EDF shown in Fig.~\ref{Fig4:BLam_A48}(b), having the largest CSB%
\begin{landscape}
\begin{table}[H]
\centering
\caption{\label{tab:E_diff} The single-$\Lambda$ binding energies $B_{\Lambda}$ as well as the difference $\Delta B_{\Lambda}$ in mirror hypernuclei with the same total isospin $T$ and opposite the third components $T_{z}$ calculated by PK1 EDF with or without $N\Lambda$ CSB interaction. The contributions from the Coulomb interaction for $B_\Lambda$ and $\Delta B_{\Lambda}$ are also listed. For comparison, available experimental data are listed. All energies are in MeV. 
\\
$^*$ The $B_\Lambda$ value, $5.064 \pm 0.332$ MeV, for $^7_{\Lambda}$He \cite{Mainz} is taken by Refs. \cite{PRC2016Gogami,Bohm,Prakash}. It should be noted that the statics of data in Refs. \cite{Bohm,Prakash} are poor. Thus, we do not mention the data in the present paper.
}
\begin{tabular}{cccrcccccccccccccccc}
\toprule
  &&\multirow{3}{*}{$T$} &\multirow{3}{*}{$T_{z}$}&& \multicolumn{5}{c}{$B_{\Lambda}$} && \multicolumn{5}{c}{$\Delta B_{\Lambda}$} \\
               \cline{6-11}\cline{13-18}
  &&&&&\multirow{2}{*}{Expt.} &\multicolumn{2}{c}{with CSB}&&\multicolumn{2}{c}{w/o CSB}&&\multirow{2}{*}{Expt.} &\multicolumn{2}{c}{with CSB}&&\multicolumn{2}{c}{w/o CSB} \\
  \cline{7-8}\cline{10-11}\cline{14-15}\cline{17-18}
  &&&&&&total&coul.&&total&coul. &&&total&coul.&&total&coul.\\
  \midrule
\multirow{3}{*}{$A=7$}&$^{7}_{\Lambda}$He & \multirow{2}{*}{$1$} & \multirow{2}{*}{$1$} &&$5.55\pm0.10\pm 0.11$~\cite{PRC2016Gogami}
                     &\multirow{2}{*}{$5.260^{+0.179}_{-0.176}$}&\multirow{2}{*}{0.007}&&\multirow{2}{*}{5.068}&\multirow{2}{*}{0.013}&&\multirow{3}{*}{0.39}& \multirow{3}{*}{0.368$\pm 0.236$}&\multirow{3}{*}{$-0.229$}&&\multirow{3}{*}{$0.003$}&\multirow{3}{*}{$-0.264$}\\
                     & & & & & $5.064 \pm 0.332^*$ \cite{Mainz}  &&&&&&&&&&&&& \\                    
                     &$^{7}_{\Lambda}$Be & $1$ & $-1$&&  $5.16\pm 0.08$~\cite{Juric}&$4.893^{+0.158}_{-0.155}$&0.236&&5.065&0.277&&\\
\midrule
\multirow{4}{*}{$A=10$}&\multirow{2}{*}{$^{10}_{\Lambda}$Be}& \multirow{2}{*}{$1/2$} & \multirow{2}{*}{$1/2$} &&   9.11$\pm0.22$~\cite{Juric,Cantwell}& \multirow{2}{*}{$9.019^{+0.105}_{-0.102}$}&\multirow{2}{*}{0.032}&&\multirow{2}{*}{8.908}&\multirow{2}{*}{0.055}&&\multirow{4}{*}{0.22/-0.04}& \multirow{4}{*}{$0.217\pm 0.132$} &\multirow{4}{*}{-0.029}&&\multirow{4}{*}{0.015}&\multirow{4}{*}{-0.044}\\
                       &&& &&   8.60$\pm0.07\pm0.16$~\cite{Gogami}& &&&\\
                       &\multirow{2}{*}{$^{10}_{\Lambda}$B} & \multirow{2}{*}{$1/2$} &\multirow{2}{*}{$-1/2$} &&   8.89$\pm0.12$~\cite{Davis}& \multirow{2}{*}{$8.803^{+0.082}_{-0.080}$}&\multirow{2}{*}{0.061}&&\multirow{2}{*}{8.893}
                       &\multirow{2}{*}{0.099}&&\\
                       &&& &&   8.64$\pm0.1$~\cite{Hasegawa,Gogami}& &&&\\
\midrule
\multirow{3}{*}{$A=12$}&$^{12}_{\Lambda}$B & $1/2$ & $1/2$ &&  11.529$\pm0.025$~\cite{TangJLab}&$11.439^{+0.111}_{-0.108}$&0.043&&11.321&0.079&&\multirow{3}{*}{0.23$\pm$0.19}& \multirow{3}{*}{
$0.229\pm 0.136$}&\multirow{3}{*}{-0.027} &&\multirow{3}{*}{0.021}&\multirow{3}{*}{-0.045}\\
                       &\multirow{2}{*}{$^{12}_{\Lambda}$C} & \multirow{2}{*}{$1/2$} & \multirow{2}{*}{$-1/2$}&&  11.30$\pm 0.19$~\cite{Gogami,Davis}&\multirow{2}{*}{$11.209^{+0.082}_{-0.079}$}&\multirow{2}{*}{0.070}&&\multirow{2}{*}{11.300}&\multirow{2}{*}{0.124}&&\\
&&&&& $11.335 \pm 0.126$ \cite{Mainz} &&&&&&&&&&&&& \\
\midrule
\multirow{6}{*}{$A=40$}&$^{40}_{\Lambda}$K & $1/2$ & $1/2$ &&      -&$19.328^{+0.059}_{-0.058}$&0.012&&19.265&0.074&&
\multirow{2}{*}{-}& \multirow{2}{*}{$0.089\pm 0.059$}&\multirow{2}{*}{0.037} &&\multirow{2}{*}{0.024}&\multirow{2}{*}{0.030}\\
                       &$^{40}_{\Lambda}$Ca & $1/2$ &$-1/2$&&$18.7\pm1.1$~\cite{Pile}&$19.239^{+0.001}_{-0.001}$&-0.025&&19.240&0.044&&\\
                       &$^{40}_{\Lambda}$Ar & $3/2$ &$3/2$ &&     -&$19.232^{+0.138}_{-0.135}$&-0.008&&19.084&0.045&&
                       \multirow{2}{*}{-}& \multirow{2}{*}{$0.248\pm 0.158$}&\multirow{2}{*}{-0.015} &&\multirow{2}{*}{0.012}&\multirow{2}{*}{-0.036}\\
                       &$^{40}_{\Lambda}$Sc & $3/2$ & $-3/2$&&    -&$18.984^{+0.080}_{-0.078}$&0.007&&19.072&0.081&&\\
                       &$^{40}_{\Lambda}$Cl & $5/2$ &$5/2$ &&     -&$19.544^{+0.192}_{-0.190}$&0.009&&19.336&0.058&&
                       \multirow{2}{*}{-}& \multirow{2}{*}{$0.378\pm 0.232$} &\multirow{2}{*}{-0.031}&&\multirow{2}{*}{0.025}&\multirow{2}{*}{-0.068}\\
                       &$^{40}_{\Lambda}$Ti & $5/2$ & $-5/2$ &&   -&$19.166^{+0.132}_{-0.130}$&0.040&&19.311&0.126&&\\
\midrule
\multirow{4}{*}{$A=42$}&$^{42}_{\Lambda}$Ca & $1/2$ &$1/2$    &&  -&$19.519^{+0.051}_{-0.050}$&-0.003&&19.464&0.060&&
\multirow{2}{*}{-}& \multirow{2}{*}{$0.053\pm0.051$}&\multirow{2}{*}{0.014} &&\multirow{2}{*}{0.007}&\multirow{2}{*}{-0.009}\\
                       &$^{42}_{\Lambda}$Sc & $1/2$ & $-1/2$  &&  -&$19.466^{+0.009}_{-0.008}$&-0.017&&19.458&0.051&&\\
                       &$^{42}_{\Lambda}$K & $3/2$ &$3/2$     &&  -&$19.613^{+0.115}_{-0.114}$&-0.014&&19.489&0.041&&
\multirow{2}{*}{-}& \multirow{2}{*}{$0.205\pm0.127$}  &\multirow{2}{*}{-0.022}&&\multirow{2}{*}{0.019}&\multirow{2}{*}{-0.043}\\
                       &$^{42}_{\Lambda}$Ti & $3/2$ & $-3/2$  &&  -&$19.408^{+0.057}_{-0.055}$&0.008&&19.470&0.084&&\\
\midrule
\multirow{4}{*}{$A=44$}&$^{44}_{\Lambda}$Ca & $3/2$ &$3/2$    &&  -&$20.025^{+0.094}_{-0.092}$&-0.017&&19.924&0.042&&
\multirow{2}{*}{-}& \multirow{2}{*}{$0.159\pm0.098$} &\multirow{2}{*}{-0.027}&&\multirow{2}{*}{0.023}&\multirow{2}{*}{-0.046}\\
                       &$^{44}_{\Lambda}$V & $3/2$ & $-3/2$   &&  -&$19.865^{+0.033}_{-0.031}$&0.010&&19.901&0.088&&\\
                       &$^{44}_{\Lambda}$K & $5/2$ &$5/2$     && - &$20.114^{+0.158}_{-0.157}$&-0.004&&19.943&0.048&&
\multirow{2}{*}{-}& \multirow{2}{*}{$0.318\pm0.184$} &\multirow{2}{*}{-0.047}&&\multirow{2}{*}{0.040}&\multirow{2}{*}{-0.081}\\
                       &$^{44}_{\Lambda}$Cr & $5/2$ & $-5/2$   &&  -&$19.797^{+0.097}_{-0.095}$&0.043&&19.903&0.129&&\\
\midrule
\multirow{2}{*}{$A=46$}&$^{46}_{\Lambda}$Ca & $5/2$ &$5/2$    &&  -&$20.513^{+0.135}_{-0.134}$&-0.008&&20.367&0.047&&
\multirow{2}{*}{-}& \multirow{2}{*}{$0.268\pm0.134$} &\multirow{2}{*}{-0.051} &&\multirow{2}{*}{0.043}&\multirow{2}{*}{-0.083}\\
                       &$^{46}_{\Lambda}$Mn & $5/2$ & $-5/2$  &&  -&$20.246^{+0.071}_{-0.069}$&0.043&&20.324&0.130&&\\
\midrule
\multirow{2}{*}{$A=48$}&$^{48}_{\Lambda}$Ca & $7/2$ &$7/2$ &&  -&$20.986^{+0.175}_{-0.174}$&0.001&&20.795&0.053&&
\multirow{2}{*}{-}& \multirow{2}{*}{$0.378\pm0.205$} &\multirow{2}{*}{-0.079}&&\multirow{2}{*}{0.068}&\multirow{2}{*}{-0.124}\\
                       &$^{48}_{\Lambda}$Co & $7/2$ & $-7/2$ &&  -&$20.608^{+0.109}_{-0.106}$&0.080&&20.727&0.177&&\\
\bottomrule
\end{tabular}%
\end{table}%
\end{landscape}%
\hspace{-0.45cm} effect in $^{48}_{\Lambda}$K. When we include the CSB interaction, it is seen that the calculated $B_{\Lambda}$s increase with larger number of total isospin $T$. It is interesting to see that calculated $B_{\Lambda}$ of $^{48}_{\Lambda}$K is the largest among $A=48$ $\Lambda$ hypernuclei. Using $^{48}$Ca target, it is possible to produce $^{48}_{\Lambda}$K by $(e,e' K^+)$ reaction at JLab.

\section{\label{sec:summary} Summary}

We studied the CSB effect in the binding energy of hyperon in the mass region of $A=7\sim 48$ in the RMF models introducing $NN$, and $\Lambda N$ interactions. The phenomenological $\Lambda N$ CSB interaction is introduced and the strength parameter is fitted to reproduce the experimental binding energy difference between the mirror hypernuclei $^{12}_\Lambda$B and $^{12}_\Lambda$C. This model is  applied to calculate the CSB energy anomaly in $A$=7 and 10 mirror hypernuclei. 
We found that our result for $A=7$ mirror hypernuclei is consistent with
the data. On the other hand, in $A=10$ hypernuclei, due to uncertainties of the experimental data is large, it is difficult to conclude whether our results are consistent or not with experimental data. In the future at J-PARC, it is planned to measure the binding energy of $^{10}_\Lambda$Be with better accuracy~\cite{EPJ_Gogami}. It might be possible to discuss on CSB effect of $A=10$ $\Lambda$ hypernuclei with future experimental data. The model is further applied to predict the binding energy difference of $A$=40 mirror hypernuclei with the isospin $T=1/2$, $3/2$, and $5/2$ nuclei together with various hyper Ca isotopes and the corresponding mirror hypernuclei. Finally the binding energy systematics of $A=$48 isobars  of hypernuclei are predicted with/without the CSB EDF. The future experimental observations in JLab and J-PARC are desperately wanted to confirm the CSB effect in hypernuclei especially in the medium-heavy hypernuclei, $A=40 \sim$ 48.

\section*{ACKNOWLEDGMENTS}
We would like to thank Dr. T. Gogami, Prof. S. N. Nakamura and
Prof. A. Gal for valuable discussions. This work was partially supported by the Natural Science Foundation of Henan Province (Grant No. 242300421156), the National Natural Science Foundation of China (No. U2032141), ERATO-JPMJER2304,
Kakenhi-23K03378, and 20H00155. YT acknowledges support from the Basic Science Research Program of the National Research Foundation of Korea (NRF) under Grants No. RS-2024-00361003, No. RS-2021-NR060129.


\end{document}